\documentclass[aps, prl, reprint, amsmath, amssymb, longbibliography, superscriptaddress]{revtex4-1}
\usepackage{miller}
\usepackage{lipsum}
\usepackage{graphicx}
\usepackage{bm}
\usepackage{fancyref}
\usepackage[T1]{fontenc}
\usepackage{newpxtext,newpxmath} 
\usepackage{soul}
\usepackage{nicefrac}
\soulregister\ref{7}
\usepackage[colorlinks,
	linkcolor=blue,
	urlcolor=blue,
	citecolor=blue]{hyperref}
\usepackage[utf8]{inputenc}
\usepackage{textcomp}
\usepackage[version=3]{mhchem}
\usepackage{xcolor}
\usepackage{titlesec}
\usepackage{palatino} 
\titlespacing*{\section}{0pt}{1.1\baselineskip}{\baselineskip}

\usepackage{pdfpages}
\usepackage{pgffor}
\usepackage{etoolbox} 

\makeatletter
\patchcmd{\@outputpage@head}{\@ifx{\LS@rot\@undefined}{}{\LS@rot}}{}{}{}
\makeatother

\begin{document}

\title[]{Bimodal Phase Diagram of the Superfluid Density in \ce{LaAlO3 / SrTiO3}\\Revealed by an Interfacial Waveguide Resonator}
	\author{Nicola \surname{Manca}$^{\ddag}$}
	\email{manca@fisica.unige.it}
	\thanks{\\\ddag~
		Present Address: Dipartimento di Fisica, Università di Genova, Via Dodecaneso 33, 16146 Genova, Italy}
	\affiliation{Kavli Institute of Nanoscience, Delft University of Technology, P.O. Box 5046, 2600 GA Delft, The Netherlands}

	\author{Daniel \surname{Bothner}}
	\affiliation{Kavli Institute of Nanoscience, Delft University of Technology, P.O. Box 5046, 2600 GA Delft, The Netherlands}
		
	\author{Ana M. R. V. L. \surname{Monteiro}}
	\affiliation{Kavli Institute of Nanoscience, Delft University of Technology, P.O. Box 5046, 2600 GA Delft, The Netherlands}
	
	\author{Dejan \surname{Davidovikj}}
	\affiliation{Kavli Institute of Nanoscience, Delft University of Technology, P.O. Box 5046, 2600 GA Delft, The Netherlands}
	
	\author{Yildiz G. \surname{Sa\u{g}lam}}
	\affiliation{Kavli Institute of Nanoscience, Delft University of Technology, P.O. Box 5046, 2600 GA Delft, The Netherlands}

	\author{Mark \surname{Jenkins}}
	\affiliation{Kavli Institute of Nanoscience, Delft University of Technology, P.O. Box 5046, 2600 GA Delft, The Netherlands}

        \author{Marc \surname{Gabay}}
        \affiliation{Laboratoire de Physique des Solides, Universite Paris-Sud and CNRS, Batiment 510, 91450 Orsay, France}
        
	\author{Gary A. \surname{Steele}}
	\affiliation{Kavli Institute of Nanoscience, Delft University of Technology, P.O. Box 5046, 2600 GA Delft, The Netherlands}
	
	\author{Andrea D. \surname{Caviglia}}
	\affiliation{Kavli Institute of Nanoscience, Delft University of Technology, P.O. Box 5046, 2600 GA Delft, The Netherlands}
	
	
	\begin{abstract}
        We explore the superconducting phase diagram of the two-dimensional electron system at the \ce{LaAlO3 / SrTiO3} interface by monitoring the frequencies of the cavity modes of a coplanar waveguide resonator fabricated in the interface itself.
	We determine the phase diagram of the superconducting transition as a function of the temperature and electrostatic gating, finding that both the superfluid density and the transition temperature follow a dome shape but that the two are not monotonically related.
        The ground state of this two-dimensional electron system is interpreted as a Josephson junction array, where a transition from long- to short-range order occurs as a function of the electronic doping.
	The synergy between correlated oxides and superconducting circuits is revealed to be a promising route to investigate these exotic compounds, complementary to standard magnetotransport measurements. 
      \end{abstract}
		
	\maketitle
The interface between the two wide band-gap insulators \ce{LaAlO3} (LAO) and \ce{SrTiO3} (STO) hosts a two-dimensional electron system (2DES)~\cite{Ohtomo2004,Cen2009} that shows superconductivity~\cite{Reyren2007} together with strong spin-orbit coupling \cite{BenShalom2010,Diez2015}, localized magnetic moments \cite{Bert2011,Li2011}, and long-range spin coherence~\cite{Ohshima2017}.
Its low carrier concentration makes this 2DES particularly sensitive to elecrostatic gating, and a superconductor-insulator transition, tunable Rashba splitting, and tunable superconductivity have already been demonstrated~\cite{Caviglia2008a, Bell2009, Caviglia2010}.
Despite being considered the prototypical correlated 2DES and a platform to realize tunable superconducting devices~\cite{Monteiro2016,Goswami2016,Thierschmann2018}, the nature of its superconducting ground state and dome-shaped phase diagram, observed upon electrostatic doping, is still not understood. This partly stems from the fact that standard magnetotransport measurements do not allow us to directly probe the superfluid density, and so far few approaches have been proposed to overcome such limitation.
Bert~\textit{et al.} employed a scanning-SQUID technique to measure the penetration depth of the screening supercurrents and extract the superfluid density of the 2DES~\cite{Bert2012}, while, more recently, Singh~\textit{et al.} calculated the variations of the superfluid density from the resonance frequency of a RLC circuit containing a lumped LAO/STO element by means of an equivalent-circuit model~\cite{Singh2018}.
A powerful tool to probe the superfluid density are the superconducting coplanar waveguide resonators (SCWRs).
SCWRs are cavities for the electromagnetic field where the frequencies of the standing waves are determined by the interplay between the geometry and electromagnetic environment~\cite{Lancaster1998, Tiggelman2009, Ahmed2015}.
The formation of quasiparticles in a SCWR causes a downshift of their resonance frequencies because the lower superfluid density increases the kinetic inductance~\cite{Baselmans2007, Vercruyssen2011}.
The high sensitivity of such devices warranted their integration in highly demanding applications, such as quantum technologies~\cite{Nataf2011, Xiang2012, Chow2014} and astronomy~\cite{Baselmans2008, Bueno2017}.

Here, we realize a SCWR by direct patterning of the 2DES at the LAO/STO interface.
This approach gives a stronger modulation of the resonance frequency in comparison to lumped-element designs used in previous implementations.
The resonance frequencies of the SCWR cavity modes are monitored as a function of the temperature and electrostatic doping and translated into variations of the Pearl length and superfluid density.
We find that both the critical temperature and superfluid density display a domelike shape as a function of the back-gate voltage, which are not monotonically related.
This nonmonotonic behavior arises from the ground state of the system, which is identified as a Josephson junction array shifting from short- to long-range order while driving the system from the under- to overdoped condition.

\begin{figure}[]
	\includegraphics[width=\linewidth]{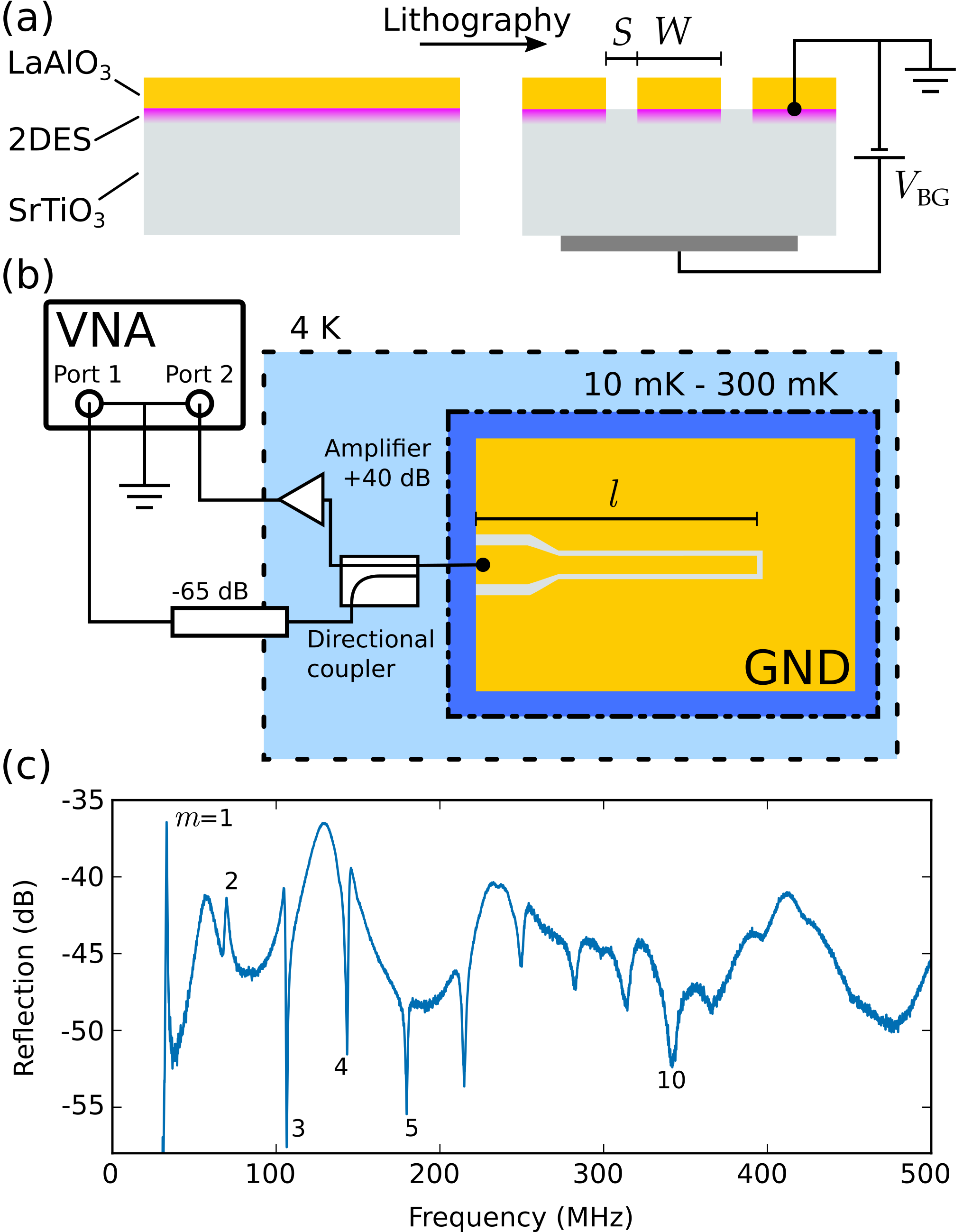}
	\caption{\label{fig:Setup}
		A coplanar waveguide resonator at the LAO/STO interface.
		(a) Sketch of the 2DES (magenta) at the LAO/STO (yellow/gray) interface before (left) and after (right) the lithography. The field effect is obtained by tuning the voltage ($V_\mathrm{BG}$) of the back gate.
		(b) Experimental setup to probe the SCWR. The $-$65~dBm attenuation is distributed across the stages from room temperature to the mK plate.
		(c) Reflection spectrum of the SCWR measured at 11~mK, $V\mathrm{_{BG}}=0$\,V, and $-95$\,dBm.
	}
\end{figure}

The pristine heterostructure is obtained by the pulsed laser deposition of 12 unit cells of crystalline LAO on top of a \ce{TiO2}-terminated \ce{SrTiO3(100)} substrate. 
As sketched in Fig.~\ref{fig:Setup}(a), by means of electron beam lithography and ion milling, we separate the central line from the ground plane (GND). The line has a width $W$ of 40~\textmu m, a nominal length $l$ of 2.5~mm, and a spacing $S$ of 10~\textmu m from the GND.
Details of the fabrication are reported in the Supplemental Material, Sec.~1.
One end of the SCWR is isolated from the GND while the other is wire bonded to the feed line.
The large impedance mismatch at the two ends of the line realizes a nearly half-wavelength ($\lambda /2$) resonator.
The cavity modes of the SCWR are measured with a vector network analyzer (VNA), as indicated in Fig.~\ref{fig:Setup}(b).
The power of the signal injected into the SCWR is about $-95$~dBm.
The sample is glued to an isolated holder with conductive silver paint, which enables field effect measurements in the back-gate geometry.
The GND and the line share the same electrical ground, so the back-gate voltage ($V\mathrm{_{BG}}$) affects both.
Figure~\ref{fig:Setup}(c) shows the response spectrum of the SCWR at $T=11$\,mK and $V\mathrm{_{BG}}=0$\,V. More than ten modes can be identified on top of an oscillating background, which stems from interference in the connecting circuitry.
We based our analysis on the modes from $m=2$ to $m=5$, because $m=1$ is at the edge of the cutoff frequency of the amplifier (30~MHz) (cf. Fig.~\ref{fig:Setup}(b)), and the higher modes show lower visibility in the explored space of parameters.
The mode visibility as a function of the temperature and $V\mathrm{_{BG}}$ is determined by the proximity to the critical coupling condition and by the damping coming from different sources of losses, with coupling, quasiparticles and dielectric losses being the major ones.
Here, the broadening and weakening of the peaks at higher frequencies are in agreement with what is expected from the \ce{SrTiO3} substrate~\cite{Davidovikj2017}.

An open-end half-wavelength resonator excited at the eigenfrequency $f_m$ can be modeled as a parallel RLC circuit with the resonance frequency
$f_m = {1}/ ({2 \pi \sqrt{L_m C}})$,
where the inductance $L_m$ is mode dependent~\cite{Goppl2008}.
In general, the inductance of a superconducting resonator is given by both a geometric and a kinetic contribution~\cite{Watanabe1994}. In our SCWR, already the first mode has a total geometric inductance of about 0.16~nH, while the kinetic inductance at $T=11$\,mK (the lowest value) is about 4~nH. $L_m$ is thus dominated by the kinetic contribution, similarly to what has been observed in other LAO/STO superconducting devices~\cite{Copie2009,Goswami2016}.
This allows us to write the two simple following expressions for the Pearl length $\Lambda$ and the 2D superfluid density $n$:
\begin{align}
\label{eq:Pearl}
\Lambda &= \alpha \left( {m}/{f_m} \right)^2,\\
\label{eq:n_2D}
n &= \beta \left( {f_m}/{m}\right)^2,
\end{align}
where $\alpha$ and $\beta$ are determined by the line geometry, the dielectric environment and the effective mass of the charge carriers, as discussed in Supplemental Material, Sec.~2.
By monitoring the magnitude of $f_m$ as a function of $T$ and $V_\mathrm{BG}$, it is thus possible to investigate their effect on the superconducting state of the 2DES.
The high sensitivity of this probing technique is due to the low density of Cooper pairs and the consequent high kinetic inductance of the 2DES.
However, we note that the calculated values of the $n$ suffer from two main sources of error.
First, the two ends of the line are not perfect mirrors and the values of $f_m$ are thus influenced by both the geometry of the launcher and the capacitance of the bonding pad.
Second, the size of the bonding wires connecting the line leads to an estimation of its effective length of 2.38$\pm$0.12~mm (see Supplemental Material, Sec.~1). We thus consider a confidence interval of $\pm10\%$ for the calculated absolute values of $n$ and $\Lambda$.

\begin{figure}[t]
	\includegraphics[width=\linewidth]{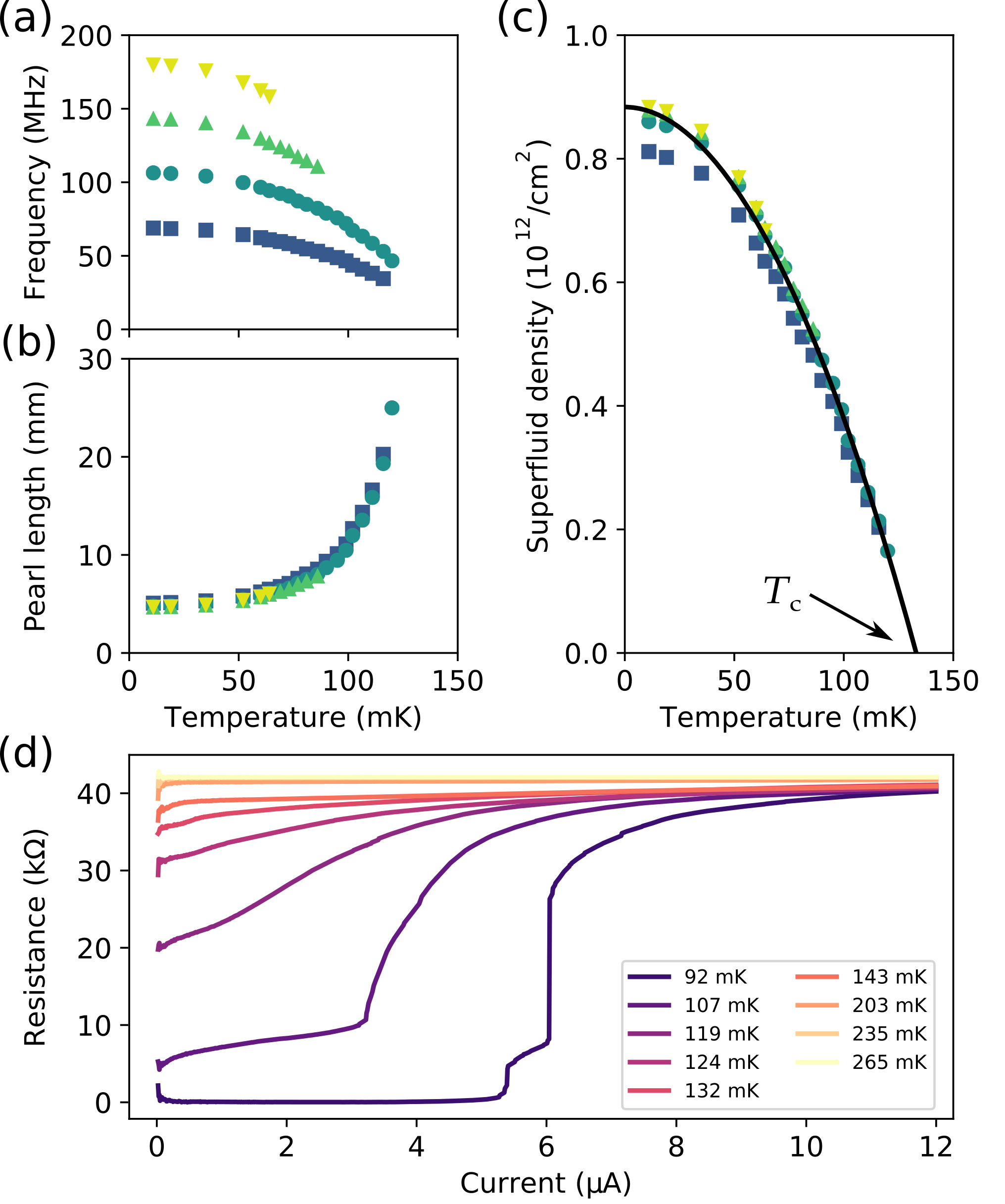}
	\caption{\label{fig:Temp_dep}
		Temperature dependence of the superfluid characteristics.
		(a) Frequencies of the cavity modes from $m=2$ (blue squares) to $m=5$ (yellow reversed triangles) at $V_\mathrm{BG}=0$\,V.
		(b) Pearl length and (c) superfluid density calculated from the data in (a) using Eqs.~(\ref{eq:Pearl}) and (\ref{eq:n_2D}).
		The solid line in (c) is the best fit of Eq.~(\ref{eq:fit_func}) for $m=3$, which gives $T_{\mathrm{c}}=133\pm2$\,mK.
                (d) Resistance-current characteristics measured on the ground plane.
	}
\end{figure}

Figure~\ref{fig:Temp_dep}(a) shows the temperature dependence of the cavity modes of the SCWR from $m=2$ to $m=5$ at $V\mathrm{_{BG}}=0$\,V.
All the modes are evenly spaced and their relative variations are in good agreement.
The disappearance of the cavity modes above 120~mK comes from the increased power dissipation associated with the formation of quasi-particles while approaching the superconducting transition.
In the normal state this device shows no resonance peaks, owing to the high resistivity of the metallic LAO/STO.
From Eqs.~(\ref{eq:Pearl}) and (\ref{eq:n_2D}) it is possible to extract the temperature dependence of $\Lambda$ and $n$ from the frequencies of each cavity mode. The calculated values are plotted in Figs.~\ref{fig:Temp_dep}(b) and \ref{fig:Temp_dep}(c), respectively.
At base temperature we have $\Lambda(11~\mathrm{mK})=4.8\pm0.5$\,mm, in good agreement to what was estimated in Ref.~\onlinecite{Goswami2016}, that increases to above 23~mm at 120~mK.
An opposite trend is observed for $n$, which starts from 0.87${\times}$10$^{12}$~cm$^{-2}$ at 11~mK with a negative slope that becomes progressively more pronounced.
We fit the temperature dependence of $n$ with a phenomenological BCS model
\begin{equation}
  \label{eq:fit_func}
  n = n_0 \left[1 - \left(\frac{T}{T_\mathrm{c}}\right)^\gamma\right],
\end{equation}
where $n_0$ is the the zero-temperature superfluid density, $T_\mathrm{c}$ is the superconducting critical temperature and $\gamma$ is an exponent which describes the opening of the gap below $T_\mathrm{c}$~\cite{Prozorov2006,Bert2012}.
The black solid line in Fig.~\ref{fig:Temp_dep}(c) is the best fit of Eq.~(\ref{eq:fit_func}) calculated for the third mode (circles in Fig.~\ref{fig:Temp_dep}).
If we consider both the second and third mode, which show the best visibility in temperature, we obtain $\gamma=1.95\pm0.23$ and $\gamma=1.96\pm0.20$, respectively.
These results are in fairly good agreement with a clean s-wave BCS scenario, where a value of $\gamma=2$ is predicted~\cite{Prozorov2006}. This is in contrast with previous works reporting $\gamma=2.8$~\cite{Bert2012} and a possible indication of lower disorder in our sample~\cite{Carbotte1990}.
The calculated critical temperature is $T_{\rm{c}}=133$\,mK, that we can compare with the transport measurements of Fig~\ref{fig:Temp_dep}(d) performed by wire-bonding the ground plane  (see also the Supplementary Material, Sec.~2).
The electrical resistance is current-dependent below 132 mK and a sharp transition is observed below 119~mK. Although a quantitative analysis is not possible because of the inhomogeneous current flow, we can consider the $T_{\rm{c}}$ as the temperature at which the electrical resistance at zero bias reaches half of its normal state, obtaining $\approx$120~mK, in good agreement with the temperature dependence of the SCWR cavity modes.

\begin{figure}[t]
	\includegraphics[width=\linewidth]{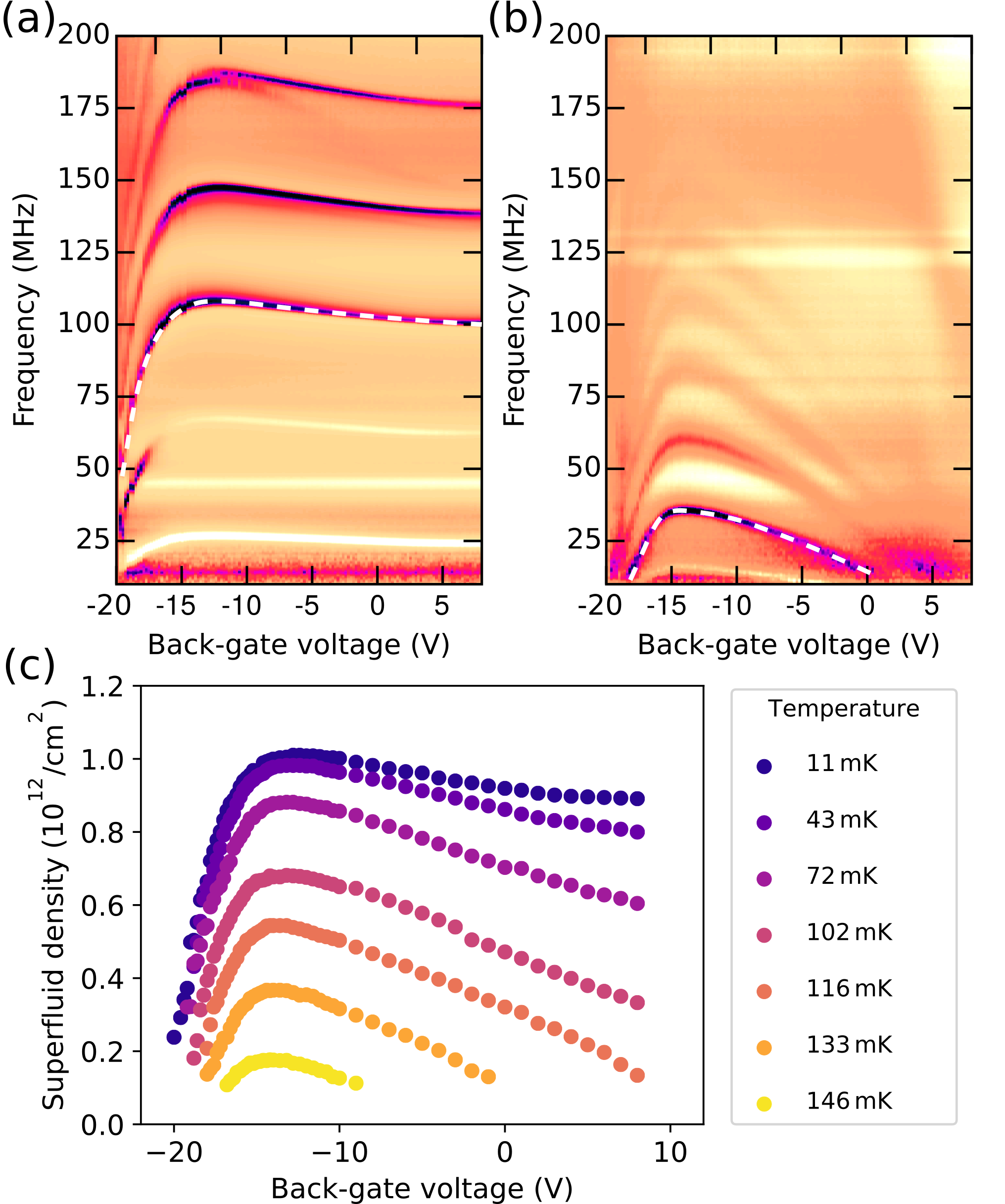}
	\caption{\label{fig:Volt_dep}
		Superfluid density under field effect.
		(a)--(b) Reflection amplitude of the SCWR as a function of $V\mathrm{_{BG}}$ at (a)~11~mK and (b)~133~mK. The white dashed line is a guide to the eye highlighting $m=3$.
		(c) Temperature-back-gate phase diagram of the superfluid density calculated from $m=3$.
	}
\end{figure}

Our experimental configuration enables tuning the superfluid density of the 2DES by electric field effect.
In Figure~\ref{fig:Volt_dep}(a) and (b) we show two colour maps of the spectral response of the SCWR measured at 11~mK and 133~mK as a function of $V\mathrm{_{BG}}$.
At the base temperature, when the gate voltage goes below $-20$~V the cavity modes rapidly shift to low frequencies and then disappear, while at positive voltages the response is rather flat with a slightly decreasing trend.
At 133~mK, instead, the superconductivity can be quenched on both ends of the phase diagram with all the modes showing a pronounced dome-like response.
This response originates from the modulation of $n$, and a possible contribution to the observed signal from the the electric-field dependence of the STO dielectric constant~\cite{Neville1972,Hemberger1995,Davidovikj2017} is discussed and ruled out in the Supplementary Material, Sec.~4.
Similarly to the analysis reported in Fig.~\ref{fig:Temp_dep}, we calculate the voltage dependence of $n$ at different temperatures using Eq.~(\ref{eq:n_2D}).
Here, we base our analysis on the third mode ($m=3$), which shows the best visibility over the whole space of parameters, while a complete dataset of the first five modes is reported in the Supplementary Material, Sec.~5.
The phase diagram of the superfluid density reported Fig.~\ref{fig:Volt_dep}(c) is characterized by a dome-like shape that becomes progressively lower and narrower for increasing temperatures.
It peaks at about $-$12.5~V, where superconductivity is still detected at 146~mK, well above $T_\mathrm{c}=132$\,mK calculated from Fig.~\ref{fig:Temp_dep}(c) and indicating that the pristine 2DES is in the over-doped regime.
The maximum value of the Cooper pairs density is $n$$\approx$$1{\times}10^{12} \mathrm{cm^{-2}}$, corresponding to an electron density of $2{\times}10^{12}~\mathrm{cm^{-2}}$.
This is in agreement with previous experiments~\cite{Bert2012,Joshua2012,Maniv2015,Singh2018}, supporting the conclusion that only a small fraction of charge carriers participate to the superconductivity.

\begin{figure}[]
	\includegraphics[width=\linewidth]{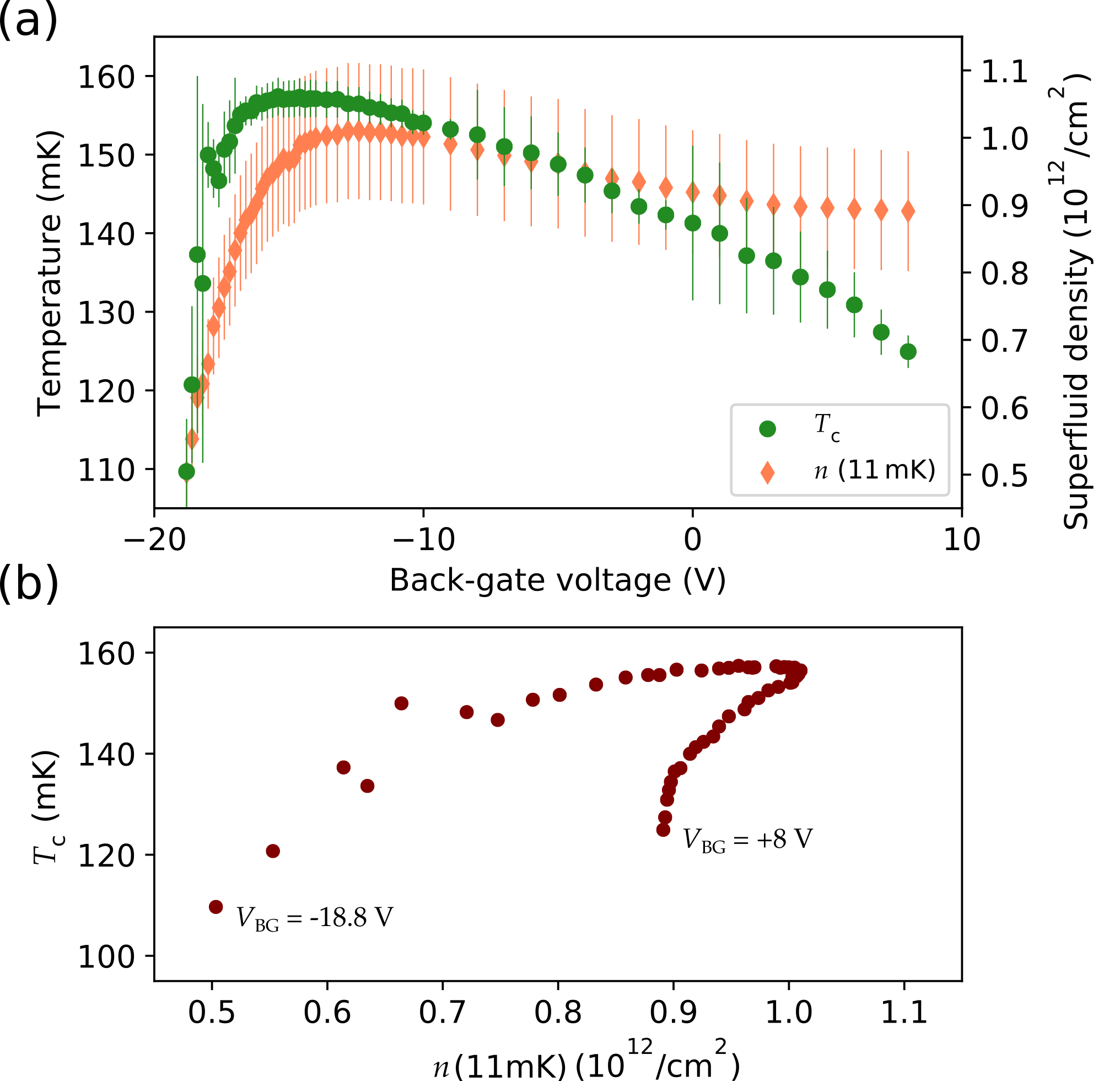}
	\caption{\label{fig:Tc_n2d}
		Analysis of the superfluid gate dependence.
		(a) $T_\mathrm{c}$ and $n(\mathrm{11~mK})$ as a function of $V_{\mathrm{BG}}$. $T_\mathrm{c}$ is calculated by fitting the data from Fig.~\ref{fig:Volt_dep}(c) with Eq.~(\ref{eq:fit_func}).
		(b) Critical temperature as a function of the superfluid density at 11~mK.
	}
\end{figure}

In Figure~\ref{fig:Tc_n2d}(a) we compare the effect of $V_{\mathrm{BG}}$ on $n$ at the base temperature and $T_{\rm{c}}$, where the latter was calculated by fitting the data reported in Fig.~\ref{fig:Volt_dep} with Eq.~(\ref{eq:fit_func}).
Both of them display a well-defined dome-shape dependence, with a maximum at $V\mathrm{_{BG}}$$\approx$$-$12.5~V.
The different position of their maximal value comes from a progressive shift of the maximum of $n(V_{\mathrm{BG}})$ with temperature and is discussed in the Supplementary Material, Sec.~6.
Different explanations have been proposed to this peculiar phase diagram. The peak of the  $T_{\rm{c}}$ has been ascribed to the Lifshitz transition, where the over-doped regime is a consequence of the onset of population of $d_{xz}$, $d_{yz}$ bands~\cite{Joshua2012}, to low-density carriers located in a high-mobility band showing non-monotonic population under field-effect due to the interplay of orbital effects and correlations~\cite{Maniv2015}, or a combination between the multi-band nature of this 2DES, electronic correlations, and disorder~\cite{Singh2018,Trevisan2018}.
The interplay between $n(11~\mathrm{mK})$ and $T_{\rm{c}}$ can be further investigated considering the $T_{\mathrm{c}}$--$n(11~\mathrm{mK})$ plot in Fig.~\ref{fig:Tc_n2d}(b).
An upper and lower branch appear, corresponding to the over- and under-doped condition and connected at $V_\mathrm{BG}\approx-$12~V (optimal doping).
A similar bimodal distribution was also reported by Bert~\textit{et al.} (grey dataset of Fig.~3 in Ref.~\cite{Bert2012}) and ascribed to inhomogeneities that locally suppress $n$ in the over-doped regime. In our case this interpretation is at variance with the results from Fig.~\ref{fig:Temp_dep}(c), where the the critical exponent $\gamma$$\approx$2 indicates low disorder.

The data presented in this work allows one to view the superconducting phase as the ground state of a Josephson junction array.
For zero gate voltage, Fig.~\ref{fig:Temp_dep}(d) shows that increasing the current $I$ in the device produces dissipation above a temperature-dependent threshold. Upon further increase of $I$, one observes a steep rise in the resistance $R$ beyond a second threshold, and $R$ ultimately levels off to its normal state value. In the Josephson junction language a BKT-like transition takes place at the lower threshold value $I_{c1}=E_J / (\varepsilon_v\Phi_0)$, where $I_{c1}$ represents the typical maximum supercurrent of a junction, $E_J$ is the Josephson coupling and $\Phi_0=h/(2e)$ is the flux quantum~\cite{Lobb1983}. The dielectric constant, $\varepsilon_v$, jumps from a finite value to infinity at the transition (strictly speaking this only holds at $T=0$\,K) and dissipation sets in above $I_{c1}$. For still larger values of the current, individual junctions in the array can sustain phase coherence (short-range order) as long as $I$<$I_{c2}=E_J/\Phi_0$. For $I>I_{c2}$, the array eventually crosses over to the normal state. In the intermediate regime, $I_{c1}<I<I_{c2}$ one may then define a bare (unrenormalized) density of Cooper pairs $n_0$
~\cite{Lobb1983} such that
\begin{align}
    \label{eq:superfluid_n}
    n_0&=\frac{2m_e}{2\pi\hbar^2}I_{c2}\Phi_0
\end{align}
From Fig.~\ref{fig:Temp_dep}(d) we determine $\frac{I_{c2}(92~\mathrm{mK})}{I_{c2}(109~\mathrm{mK})}=1.8$.
This number is in line with the ratio that one can extract from the temperature dependence of the cavity modes in Fig.~\ref{fig:Temp_dep}(c), which is $\frac{n_{0}(92~\mathrm{mK})}{n_{0}(109~\mathrm{mK})}=1.79$.
From Fig.~\ref{fig:Temp_dep}(d) we see that for increasing temperatures $I_{c1}$ and $I_{c2}$ are no more distinguishable and $T_\mathrm{c}$ goes to zero together with $n$.
The temperature dependence of $n$ reported in Fig.~\ref{fig:Temp_dep}(c) (overdoped condition) can be thus interpreted as a regime where the superconducting islands are large enough to sustain long-range coherence and $n(T)$ follows the simple BCS model of Eq.~(\ref{eq:fit_func}), that in this case was found compatible with a clean s-wave superconductor.
In the under-doped condition the lowered electron density makes the superconducting puddles to lose connection, resulting in a transition dominated by short-range order. 
This picture explains the two branches of Fig.~\ref{fig:Tc_n2d}(b), originating form the different nature of the ground state in the two regimes, in agreement with recent experimental results~\cite{Biscaras2013, Chen2018}.
We may now justify the fact that transport properties of the 2DEG in this device can be related to their counterpart in Josephson junction networks.
In Fig.~\ref{fig:Temp_dep}(d) the steep rise in the resistance at $T=92$\,mK is observed at $I_{c1}(92~\mathrm{mK})=5.4$\,\textmu A. The equations leading to Eq.~(\ref{eq:superfluid_n}) also predict that the value of the critical current per junction is $I_{c2}(92~\mathrm{mK})$$\approx$37~nA and $I_{c2}(109~\mathrm{mK})$$\approx$20~nA, suggesting that about 150 parallel channels are contributing to the electrical current.

Below $T=105$\,K, STO undergoes a structural transition from a cubic to a tetragonal phase~\cite{Unoki1967, Roy2017}. Current maps of charge flow in \ce{LAO / STO} reveals a filamentary structure of the pattern which is related to the striped electrostatic potential modulation arising from the tetragonal domains in \ce{STO} at low temperature~\cite{Frenkel2016, Frenkel2017, Honig2013, Noad2018, Pai2018}. However, the spatial resolution is not sufficient to simultaneously image the current paths and the domain boundaries; the size of the latter does not exceed 500--600~nm. Estimations of the wall (twins, dislocations) widths in the bundles reported by several authors are between few tens of nanometers~\cite{Frenkel2017,Honig2013} and 1--10~nm~\cite{Szot2006,Schiaffino2017}. If the conducting channel consists of a bundle of filaments of micrometer size, one could consider each filament as forming a junction neighbouring filaments being separated by 10~nm walls. By considering about 150 parallel junctions, one recovers an approximate size of a few microns for the bundle.

In conclusion, we studied the superfluid density at the LAO/STO interface by means of a coplanar waveguide resonator patterned into the heterostructure itself.
With no gate applied, the temperature dependence of the superfluid density is in good agreement with a clean s-wave BCS superconductor, while under field effect both the critical temperature and the superfluid density show a dome-shaped phase diagram, leading to a multi-valued relationship between them.
The comparison between transport data and cavity resonances suggests that the ground state of this 2DES is a Josephson junction array undergoing a transition between long- and short-range order under electrostatic doping.
We foresee future experiments taking advantage of the high sensitivity of this technique, as an example by combining superconducting resonators and magnetotransport measurements to explore different oxide-based 2DES.

\section*{Acknowledgements}
The research leading to these results has received funding from the European Research Council under the European Union’s H2020 programme/ERC Grant Agreement No. 677458, the project Quantox of QuantERA ERA-NET Cofund in Quantum Technologies, the Netherlands Organisation for Scientific Research (NWO/OCW) as part of the Frontiers of Nanoscience program (NanoFront), and by the Dutch Foundation for Fundamental Research on Matter (FOM)

\section*{Open Data}

The numerical data shown in figures of the manuscript and the supplemental material can be donwloaded from the Zenodo online repository: ~\href{http://dx.doi.org/10.5281/zenodo.2530003}{http://dx.doi.org/10.5281/zenodo.2530003}

\bibliography{library}
\bibliographystyle{apsrev4-1}

\newpage\newpage



\section*{Supplementary material}

# Bimodal Phase Diagram of the Superfluid Density in LaAlO$_3$/SrTiO$_3$ Revealed by an Interfacial Waveguide Resonator

## *Supplementary Material*

Nicola Manca,[1,*] Daniel Bothner,[1] Ana M. R. V. L. Monteiro,[1] Dejan Davidovikj,[1] Yildiz G. Sağlam,[1] Mark Jenkins,[1] Marc Gabay,[2] Gary Steele,[1] and Andrea D. Caviglia[1]

[1]*Kavli Institute of Nanoscience, Delft University of Technology,*
*P.O. Box 5046, 2600 GA Delft, The Netherlands*

[2]*Laboratoire de Physique des Solides,*
*Universite Paris-Sud and CNRS, Batiment 510, 91450 Orsay, France*

(Dated: January 2, 2019)

* manca@fisica.unige.it

This supplemental material contains the following:

Supplementary Sec. 1: Details of the device fabrication.

Supplementary Sec. 2: Calculation of the Pearl length and superfluid density from $f_m$.

Supplementary Sec. 3: DC V–I measurements.

Supplementary Sec. 4: Contribution of the field-dependent dielectric constant of $SrTiO_3$.

Supplementary Sec. 5: Voltage and temperature dependence of the first five cavity modes.

Supplementary Sec. 6: Temperature dependence of the optimal doping condition.

**Sec. 1.   Details of the device fabrication**

The LAO film was deposited by means of Pulsed Laser Deposition (PLD) at 840 °C in $6\times10^{-5}$ mbar of pure oxygen on tiop of a syngle-crystal SrTiO3(001) substrate. The laser pulses were supplied by a KrF excimer source ($\lambda$ = 248 nm) with an energy density of 1 J cm$^{-2}$ and a frequency of 1 Hz. In order to remove oxygen vacancies, the growth process was followed by an annealing in 300 mbar of oxygen, at 600 °C for 1 hour. The sample was subsequently cooled down to room temperature at a rate of 10 °C min$^{-1}$ in the same oxygen atmosphere. The growth process was monitored in-situ using reflection high-energy electron diffraction (RHEED), which indicated a layer-by-layer growth mode. The SCWR is fabricated by covering the sample with a protective layer of polymethyl-methacrylate (PMMA) which is patterned by Electron Beam Lithography (EBL). After the developing process, the exposed areas are etched by Ar-milling (500 eV, 0.2 mA cm$^{-2}$) to remove the LAO layer and obtain electrical isolation between the central line and the ground plane. For the data analysis it is important to evaluate the actual length of the SCPW. We do this by considering Fig. SF 1, which shows a picture of the final device connected to the feed-line by several bonding wires. The error on the line length is evaluated by considering the size of the contact area, giving $l$=2.38±0.12 mm.

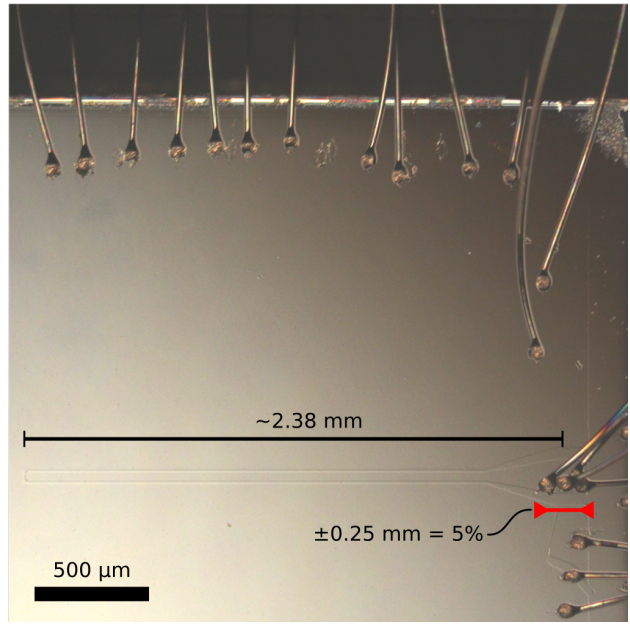

FIG. SF 1.   Picture of the final device.

## Sec. 2. Calculation of the Pearl length and superfluid density from $f_m$

The most important quantities to determine the frequencies of the cavity modes in a low-loss transmission line having a given geometry are the capacitance $C'$ and inductance $L'$ per unit length. For a coplanar waveguide with large ground planes these parameters can be calculated from the line width $W$ and the spacing toward the ground plane $W$. The high ratio between the thickness of the STO substrate ($t$=500 µm) and the cross-sectional dimensions of the SCWR ($W$ and $S$) allows to consider our device in the thick-substrate limit ($t \gg W + 2S$), the geometric inductance and capacitance can be thus calculated as

$$C' = 4\varepsilon_0 \varepsilon_{\text{eff}} \frac{K(k_0)}{K(k'_0)} \quad , \quad L'_g = \frac{\mu_0}{4} \frac{K(k'_0)}{K(k_0)} \tag{SE 1}$$

where

$$k_0 = \frac{W}{W + 2S} \quad , \quad k'_0 = \sqrt{1 - k_0^2} \quad , \quad \varepsilon_{\text{eff}} = \frac{\varepsilon_r - 1}{2} \quad ,$$

$K(x)$ is the complete elliptic integral of the first kind, $\varepsilon_0$ is the vacuum permittivity, $\varepsilon_r$ is the dielectric constant of the substrate and $\mu_0$ is the vacuum permeability. An open-ended transmission line, excited at the resonance frequency $f_m$ of the $m$-th mode, can be modelled as a parallel LC circuit with a capacitance $C$ and a mode-dependent inductance $L_m$.

$$C = \frac{C'l}{2} \quad ; \quad L_m = \frac{2L'l}{m^2 \pi^2} \quad ; \quad f_m = \frac{1}{2\pi \sqrt{L_m C}}. \tag{SE 2}$$

Typically, the total inductance per unit length $L'$ corresponds to the geometric contribution ($L' \approx L'_g$). However, as discussed in the main text, in the 2DES at the STO/LAO interface the kinetic inductance can be identified in good approximation with the total inductance per unit length ($L' \approx L'_k$). In Ref. S1, the Pearl length $\Lambda$ is estimated to be of the order of several mm, which is much larger than the width $W$ of the SCWR. In the $\Lambda \gg W$ limit, $\Lambda$ is related to $L'_k$ by the following simple expression [2]

$$\Lambda = 2W \frac{L'_k}{\mu_0}. \tag{SE 3}$$

Furthermore, since the thickness of the superconducting layer is small compared to the magnetic penetration depth $\lambda_L$ ($d \approx 10$ nm $\ll \lambda_L$) [3], it is possible to rewrite Eq. (SE 3) as a function of the superfluid density of STO/LAO with the inductance of the LS-CPW:

$$\Lambda = \frac{2\lambda_L^2}{d} \quad ; \quad \lambda_L = \frac{1}{q}\sqrt{\frac{m_e}{\mu_0 n}} \quad ; \quad L'_k = \frac{\mu_0 \Lambda}{2W} \rightarrow L'_k = \frac{1}{dWq^2} \frac{m_e}{n}, \tag{SE 4}$$

where $n$ the planar density of Cooper pairs, $m_e$ their effective mass and $q = 2e$ their charge. $m_e$ is evaluated as two times the electron effective mass in LAO/STO $m = 2m^* = 2 \cdot 1.46\, m_e$ and $m_e$ is the the bare electron mass [4]. By combining Eq. SE 1, SE 2 and SE 4 we can write a direct relationship between the resonance frequencies of the modes of the SCWR and the superfluid density

$$f_m = \frac{m}{2l}\sqrt{\frac{K(k_0')}{K(k_0)}\frac{Wq^2}{4m_e \varepsilon_0 \varepsilon_{\text{eff}}}}\sqrt{n} \tag{SE 5}$$

This relationship provides the $\alpha$ and $\beta$ parameters introduced in the main text, which are defined as

$$\alpha = \frac{W}{8dl^2}\frac{K(k_0')}{K(k_0)}\frac{1}{\mu_0 \varepsilon_0 \varepsilon_{\text{eff}}} \tag{SE 6}$$

$$\beta = \frac{16l^2}{W}\frac{K(k_0)}{K(k_0')}\frac{\varepsilon_0 \varepsilon_{\text{eff}}}{q^2} m_e \tag{SE 7}$$

5

**Sec. 3. DC V–I measurements**

In Figure SF 2 we show a series of voltage vs current curves measured at different temperatures at zero back-gate voltage. The data reported in Figure 2(d) in the main text were calculated as the ratio between measured voltage and applied current. These data were acquired by wire-bonding the ground plane of the LAO-STO sample with four aligned contacts, as schematically shown in the panel (b) of the figure.

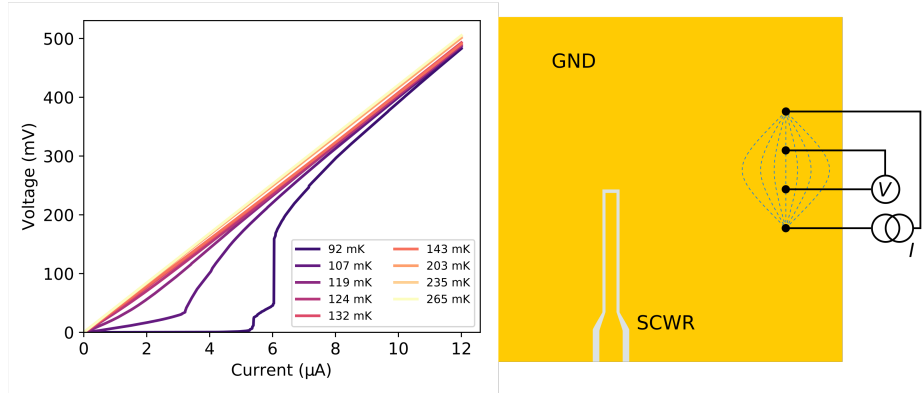

FIG. SF 2. (a) Voltage-current relationship as a function of temperature measured by wire bonding the ground plane. (b) Schematic drawing of the measurement configuration. The blue dashed lines indicates the current flow in the ground plane (GND) surrounding the superconducting coplanar waveguide resonato (SCWR).

**Sec. 4. Contribution of the field-dependent dielectric constant of SrTiO$_3$**

A different explanation for the observed shift of the cavity modes as a function of the backgate voltage reported in Fig. 3(a) and (b) in the main text may involve the electric-field dependence of the dielectric constant of the SrTiO$_3$ substrate. This would lead to a variation in the resonance frequencies of the SCWR due to a different capacitance instead of kinetic inductance. It was recently reported high tunability (>200%) in CPWs fabricated on top of SrTiO$_3$ substrate upon the application of small voltages (below 15 V) [5]. This result was based on the tuning of the dielectric constant of the STO between the line and the ground plane by field-effect. However, in our case he maximum back-gate voltage (−20 V) corresponds to an electric field of about 0.4kV/cm, which is too small to be the origin of the observed frequency shift [6]. To confirm this we fabricated a superconducting Molybdenum-Rhenium coplanar waveguide on top of a 5×5 mm$^2$ SrTiO$_3$(001) substrate and applied a field effect in the back-gate geometry. In this case the superfluid density is much higher than the 2DES in LAO/STO, and response from $V_{BG}$ would mean that the substrate contribution is not negligible. As shown in Fig. SF 3, upon the application of a back-gate voltage of 20 V a barely visible frequency shift of the cavity modes was detected, meaning that the contribution from the field-dependent dielectric constant of the substrate can be neglected.

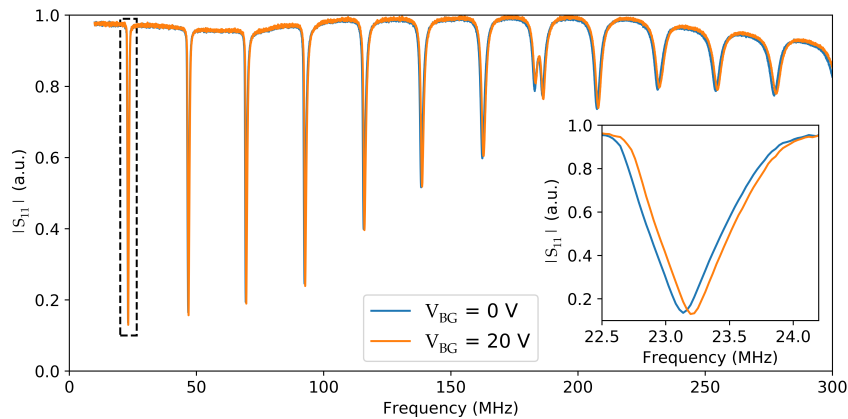

FIG. SF 3. Spectrum showing the cavity modes of a superconducting Molybdenum-Rhenium coplanar waveguide patterned on top of a 5×5 mm$^2$ substrate of SrTiO$_3$ for $V_{BG}$=0 V and 20 V. The inset shows a magnification of the region inside the dashed line (first mode).

**Sec. 5. Voltage and temperature dependence of the first five cavity modes**

Figure SF 4 shows the value of the resonance frequencies of the cavity modes from $m=1$ to $m=5$ extracted from the raw data, as the ones presented in Figures 3(a) and 3(b) in the main text. The first mode lies below the cut-off frequency of the amplifier (see Figure 1(b) in the main text) and is thus affected by a systematic error. All the modes show a similar dependence from back-gate voltage and temperature, however the third mode has the best visibility in the whole space of parameters and was employed to perform the analysis reported in Figure 3(c) and Figure 4 of the main text.

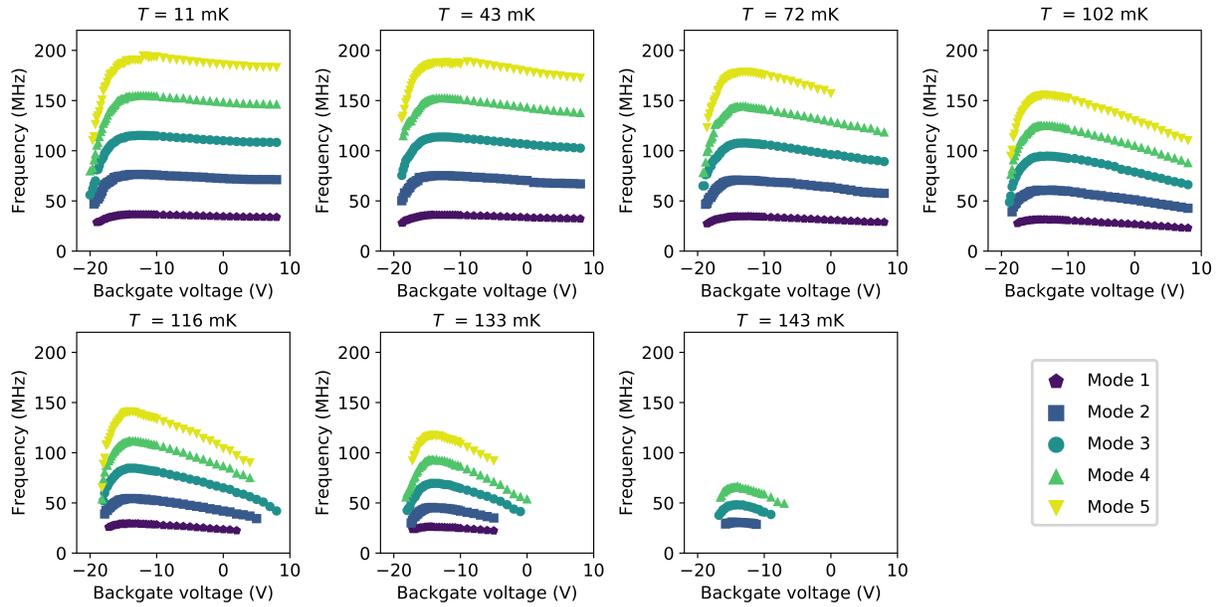

FIG. SF 4. Resonance frequencies of the five lowest cavity modes as a function of $V_{\text{BG}}$ and $T$.

**Sec. 6. Temperature dependence of the optimal doping condition**

In Figure SF 5 we show the interplay between temperature and back-gate voltage in determining the optimal doping condition, which is evaluated from the data reported in Fig. 3(c) of the main text. The optimal doping is defined as the maximum value of $n$ as a function of $V_{BG}$ for a given temperature. The value of the back-gate voltage corresponding to the optimal doping condition shifts downward with increasing temperature, similarly to what the maximal superfluid density does. Such a temperature dependence is typical when considering $n$, but it is not expected when considering $V_{BG}$, at least within this range of temperatures where the dielectric properties of the SrTiO$_3$ substrate are constant. The purple star marker indicates the maximal critical temperature extracted from Fig. 4(a) in the main text, i.e. the fits at fixed voltage of the data reported in Fig. 3(c) and Supplementary Material, Sec. 5. This shows a good agreement between the voltage sweeps at fixed temperature and the critical temperatures extracted from the fits of the data.

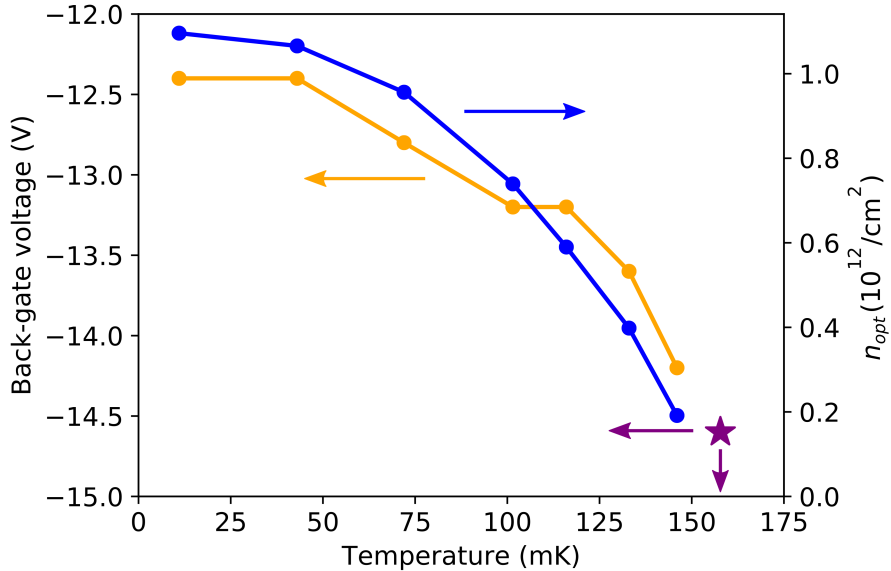

FIG. SF 5. Comparison between the temperature dependence of the value of $V_{BG}$ maximizing $n$ (orange line) and the corresponding $n$ (blue line) extracted from the data of Fig. 3. The purple star indicates the highest $T_c$, extracted from Fig. 4(a), and the corresponding $V_{BG}$.

9



\end{document}